\documentclass[prl,twocolumn,reprint]{revtex4-1}
\usepackage{amsmath}
\usepackage{graphicx}
\usepackage{subcaption}
\newcommand{\vphi}{\varphi}

\newcommand{\rhov}{\rho_{\rm v}}
\newcommand{\rhos}{\rho_{\rm s}}
\newcommand{\Cins}{C_{\rm ins}}
\newcommand{\Lsf}{L_{\rm sf}}
\newcommand{\bea}{\begin{eqnarray}}
\newcommand{\eea}{\end{eqnarray}}
\newcommand{\be}{\begin{equation}}
\newcommand{\ee}{\end{equation}}
\newcommand{\Real}{\mbox{Re}}
\newcommand{\Imag}{\mbox{Im}}
\def\tsigma{{\tilde \sigma}}
\def\rmv{{\rm v}}
\graphicspath{{./}}
\begin{document}
\title{Critical capacitance and charge-vortex duality near the superfluid to insulator transition}

\author{Snir Gazit}
\author{Daniel Podolsky}
\author{Assa Auerbach}
\affiliation{Physics Department, Technion, 32000 Haifa, Israel}

\date{\today}
\begin{abstract}
Using a generalized reciprocity relation between  charge and vortex conductivities at  complex frequencies in two space dimensions, we identify the capacitance in the insulating phase as a measure of vortex condensate stiffness.
We compute the ratio of boson  superfluid stiffness to vortex condensate stiffness at mirror points  to be $  0.21(1) $ for the relativistic O(2) model.
The product of  dynamical conductivities at mirror points is used as a quantitative measure of deviations from self-duality between charge and vortex theories. We propose the finite wave vector compressibility as an experimental measure of the vortex condensate stiffness for neutral lattice bosons.
\end{abstract} 
\maketitle

Two dimensional superfluid to insulator transitions  (SIT)  have been observed in diverse systems: {\em e.g.} Josephson junction arrays \cite{SIT_Josephson}, cold atoms trapped in  optical lattices \cite{Esslinger_2DSFMI,Spielman_2DSFMI,endres}, and disordered superconducting films \cite{Goldman_SIT_ThinFilms}.    Recent experiments have uncovered important {\em dynamical} properties near the quantum critical point:  A softening amplitude (Higgs) mode  observed in the optical lattice \cite{endres}, and   a critically suppressed threshold frequency seen by terahertz conductivity  in superconducting films \cite{shermanThz}. 
These have motivated numerical studies of real-time correlations near criticality\cite{Gazit_Dynamics,swanson2013dynamical,Rancon_higgs}, and 
novel ideas from holography \cite{Witczak_Cond,Kun_cond,holo_sachdev}.

Three decades ago, Fisher and Lee \cite{FisherLee}  showed that the SIT can be described as  a {\em Bose condensation} of quantum vortices.  Despite the appeal of this description, $\rhov$, the {\em vortex condensate stiffness},  has remained an elusive observable, for which no experimental probe has yet been proposed. Also, to our knowledge, $\rhov$ has not been calculated near the critical point, for any microscopic model.

In this Letter, we address this problem, by using
an exact reciprocity relation between complex  {\em dynamical} conductivities of bosons  ($\sigma $) and vortices ($\sigma_{\rm v}$):
\be
\sigma (\omega) \times \sigma_{\rm v}(\omega) = { q^2 / h^2 }\,,
\label{eq:recip}
\ee
where $q$ is the boson charge ($=2e$ in superconductors) \cite{fisher_UniCond}. At low frequencies, this equation is dominated by the {\em reactive} (imaginary) conductivities.  The superfluid stiffness $\rhos$ in the superfluid phase can be measured by the low frequency inductance $\Lsf$, $\rhos=\hbar/(2\pi \sigma_q\Lsf)$, where $\sigma_q= q^2/h$ is the quantum of conductance.   Equation (\ref{eq:recip}) allows us to identify the elusive vortex condensate stiffness  with the  {\em capacitance per square} in the insulating phase, $\Cins=\hbar  \sigma_q / (2\pi\rho_{\rm v})$. 

The charge-vortex duality (CVD) is an {\em exact} mapping between boson and vortex degrees of freedom near the SIT. In the presence of particle hole symmetry, the CVD is mathematically equivalent to the well known classical statistical mechanics duality mapping between the XY model and a lattice superconductor in three space dimensions \cite{CVD_Peskin,CVD_Dasgupta}.  

A related but distinct concept to CVD is {\em self-duality}, a property of certain physical systems in which the  original and dual degrees of freedom satisfy identical dynamics.  If the CVD mapping were self-dual then the universal critical conductivity at the SIT would equal exactly $\sigma_q$ \cite{FisherLee}. Experiments, however, have measured non-universal values of the critical conductivity\cite{Goldman_SIT_ThinFilms}, indicating that the boson and vortex theories are {\em not} self-dual. This is attributed mainly to the different interaction ranges of bosons (in the superfluid) and vortices (in the insulator). Moreover, in real experiments several additional factors can spoil self-duality: (i) potential energy (both confining and disordered), which couples differently to  charges and vortices (ii) fermionic (Bogoliubov) quasiparticles in superconductors, produce dissipation, which can alter the phase diagram from the purely bosonic theory.

\begin{figure}[b]
\centering
\includegraphics[scale=.5]{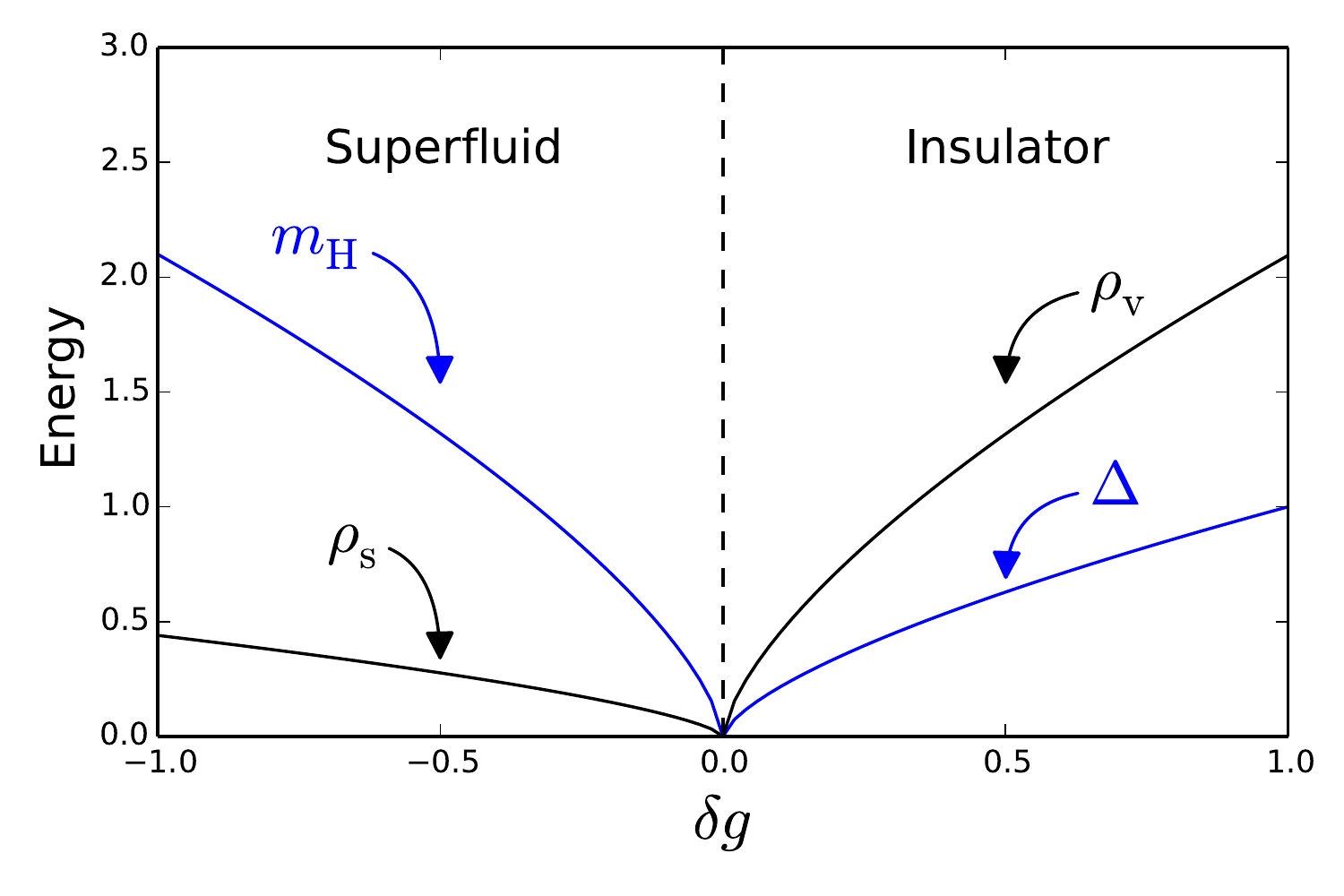}
\caption{Critical energy scales near the SIT computed by QMC. The superfluid is characterized by the  mass of the amplitude mode, $m_H$, and the superfluid stiffness, $\rhos$; the insulator by the single particle gap, $\Delta$, and the vortex condensate stiffness, $\rhov$. The amplitude ratios $m_H(-\delta g)/\Delta(\delta g)=2.1(3) $\cite{gazit_fate}, $\rhos(-\delta g)/\Delta(\delta g)=0.44(1)$ \cite{Gazit_Dynamics}, and $\rhov(\delta g)/\Delta(\delta g)=2.1(1)$ are universal.}
\label{fig:critical_energies}
\end{figure}
 
In a self-dual theory, one expects $ \rhos(-\delta g)  = \rhov (\delta g)$, where  $\pm \delta g$ are  mirror points on either side of the SIT.  Figure \ref{fig:critical_energies} depicts all the critical energy scales of the relativistic O(2) field theory, obtained by large-scale Monte Carlo simulation.  In addition to the Higgs mass $m_{\rm H}$ and the charge gap $\Delta$, which vanish  at the critical point, we compare the  energy scales  $\rhos$ and $\rhov$ 
which are also critical, but have different relative amplitudes.  The ratio $\rhos(-\delta g)/\rhov(\delta g)=0.21(1)$ differs form unity and hence quantifies the deviation from self-duality.

It is interesting to ask whether self-duality is better satisfied at finite frequencies.  To address this, we propose the product function
\be
{\cal R}(z)\equiv \sigma(z, -\delta g) \times  \sigma(z, \delta g) /\sigma_q^2\,,
\label{eq:calR}
\ee
as a measure of self-duality between mirror points. Here, $z$  denotes either a real or a Matsubara frequency.

The   high frequency conductivity \cite{foot}  (after removal of cut-off dependent corrections) reaches a universal value $\sigma^*=0.355(5)\sigma_q$ \cite{Kun_cond,Witczak_Cond}. 
We compute the function ${\cal R}(i\omega_m)$ and adress its implications to CVD.  We conclude by proposing an experimental measure of the vortex condensate stiffness $\rhov$ for neutral bosons in an optical lattice.

\emph{Vortex transport theory}--- 
Boson charge current $\vec{j}$ is driven by an electrochemical field $\vec{E}$.  Vortices are point particles in two dimensions.
The vorticity current $\vec{j}_{\rm v}(t)$ is driven by the Magnus field  $\vec{E}_{\rm v}$.
 Hydrodynamics dictate simple relations between electrochemical field and vortex number current, and between boson charge current and Magnus field \cite{Auerbach_CVD}:
 \be
E^\alpha_{v} = {h\over q} \epsilon^{\alpha\beta} j^\beta \, ,~~~E^\alpha = {h\over q} \epsilon^{\alpha\beta} j^\beta_{v}\, ,
\label{magnus}
\ee
where $\epsilon=i\sigma^y$ is the two dimensional antisymmetric tensor.
We note that Eqs.~(\ref{magnus}) are {\em instantaneous}. 
Conductivity relates currents to their driving fields,  
\be
j^\alpha_{(\rmv)} (t) = \int^t_{-\infty} dt' \sigma_{(\rmv)}^{\alpha\beta}(t-t') E_{(\rmv)}^\beta(t')\, . 
\ee
By Fourier transformation, the complex dynamical conductivities
obey a reciprocity relation
$\varepsilon^\top \sigma_\rmv\varepsilon = ( q^2 / h^2 )\sigma^{-1}$. For the case of an isotropic longitudinal conductivity $\sigma^{xx}=\sigma^{yy}=\sigma$, one obtains 
the reciprocity Eq.~\eqref{eq:recip}, which can be analytically continued to Matsubara space $\omega \to i\omega_n$. 

\emph{Model and observables--} For numerical simulations we study the discretized partition function 
$\mathcal{Z}=\int \mathcal{D} \vphi \mathcal{D} \vphi ^*e^{-\mathcal{S}[\vphi,\vphi^*]}$, where the real action $S$ on Euclidean space-time is 
\begin{equation}
 \mathcal{S}=\sum_{\langle i,j\rangle} \vphi_i\vphi_j^* + \mbox{c.c} +2\mu \sum_i |\vphi_i|^2 +4g \sum_i |\vphi_i|^4\,.
 \label{eq:lat_model}
 \end{equation}
Here  $\vphi_i$ are complex variables defined on a cubic lattice of size $L\times L\times \beta$.  We take $\beta=L$ throughout. For  $\mu<0$, this model undergoes a continuous zero temperature quantum phase transition (QCP) between a superfluid with $\langle\vphi\rangle\ne 0$ for $g<g_c$ and an insulator with $\langle\vphi\rangle= 0$ for $g>g_c$.  We define the quantum detuning parameter $\delta g=(g-g_c)/g_c$.  

The critical energy scales near the SIT, as shown in Fig~\ref{fig:critical_energies}, in the superfulid phase are the amplitude mode mass $m_H$ and the superfluid stiffness $\rhos$ \cite{dpss,gazit_fate}.  In the insulating phase excitations are gapped, with single-particle gap $\Delta$.

The lattice current field is  $J_{i,\eta}=-\frac{\delta \mathcal{S}}{\delta A_{i,i+\eta}}$, where we have introduced a $U(1)$ lattice gauge field by Peierls substitution $\vphi_i\vphi_{i+\eta}^*\to\vphi_i\vphi_{i+\eta}^*e^{iA_{i,i+\eta}}$. The
 dynamical conductivity is given by the  current-current correlation function,
\bea
\tsigma(\omega_m)&=& -\frac{\Pi_{xx}(\omega_m) }{\omega_m}\,,\nonumber\\
\Pi_{xx}(\omega_m )&=& {1 \over L^2 \beta} \sum_{i,j} e^{i \omega_m \tau_{ij}}\frac{\delta \left \langle J_{i,x} \right\rangle} {\delta A_{j,x}}\,,
\label{Kubo}
\eea
where $\omega_m=2\pi m T$ is a  Matsubara frequency and $\tau_{ij}$ is the discrete imaginary time interval between points $i,j$.  Remarkably, in $2+1$ dimensions the conductivity has zero scaling dimension \cite{fisher_UniCond}, such that it is given by a universal amplitude with scaling form
\begin{equation}
\tsigma(\omega_m)=\sigma_q \Sigma_{\pm}(\omega_m/\Delta)\,,
\label{eq:sigmascale}
\end{equation}
where $\Sigma_+$ ($\Sigma_-$) belongs to the  insulating (superfluid) phase.   
Real frequency dynamics  can be obtained by analytic continuation $\sigma(\omega)=\tsigma(\omega_m\to-i\omega+0^+)$.
 
In the superfluid phase, the reactive conductivity diverges as $\Imag \sigma_{\rm sf}(\omega)=2\pi\sigma_q \rhos(-\delta g)  /(\hbar \omega)$.  Previous calculations~\cite{Podolsky_visibility} show that  the dissipative component has a small sub-gap contribution below the Higgs mass, $0<\omega\ll m_H(-\delta g)$ which goes as $\Real (\sigma_{\rm sf}(\omega))\sim \omega^5$. This is negligible  as $\omega\to 0$ and the analytic continuation to Matsubara frequency yields
\begin{equation}
\tsigma_{\rm sf}(\omega_m) \sim {2\pi\sigma_q\, \rhos \over  \hbar \omega_m}\,,~~~ ({\mathrm{for}} ~~ \omega_m\ll m_H).
\label{eq:sigmasf}
\end{equation}

In the insulator, the dissipative conductivity vanishes identically below the charge gap $\Delta(\delta g)$~\cite{damle_UniCond,Gazit_Dynamics}. The reactive conductivity vanishes linearly with frequency 
$\Imag \sigma_{\rm ins}(\omega)= - C_{\rm ins}\omega$, where $C_{\rm ins}$ is the capacitance per square. 
As a result,  the conductivity, as function of the complex frequency $z$, has radius of convergence of $2\Delta$ about $z=0$ and in the low frequency limit, it is given by $\sigma_{\rm ins}=-i C z$. This can be analytically continued to Matsubara space by setting $z=i\omega_m$, which yields
\begin{eqnarray}
\tsigma_{\rm ins}(\omega_m) \sim  {C_{\rm ins}\, \omega_m }\,,~~~ ({\mathrm{for}} ~~ \omega_m\ll \Delta).
\label{eq:sigmains}
\end{eqnarray}
 Eq.~(\ref{eq:sigmascale}) implies that the capacitance $C_{\rm ins}$ diverges near the QCP as $C_{\rm ins}\sim 1/\Delta$.  
The capacitance measures the dielectric response of the insulator.  Its divergence reflects the large particle-hole fluctuations near the transition.

In the vortex description the insulator is a bose condensate of vortices, with a low frequency vortex conductivity $\tilde{\sigma}_\rmv(\omega_m)=\rhov/(\hbar ^2 \omega_m)$. As a consequence, $\rhov$ can be defined in terms of the capacitance by applying Eq.~(\ref{eq:recip}),
\begin{equation}
\rhov \equiv   {\hbar \sigma_q \over 2\pi C_{\rm ins}}.
\label{eq:C_rhov}
\end{equation}
We shall use this important relation to test for self-duality in the 2+1 dimensional O(2) field theory.

{\em Methods --} A large scale QMC simulation of Eq.~(\ref{eq:lat_model}) is used to evalaute Eq.~(\ref{Kubo}).  To suppress the effect of critical slowing down near the phase transition we use the classical Worm Algorithm \cite{prokofev_worm_2001}.  This method samples closed loop configurations of a dual integer current representation of the partition function. This enables us to consider large systems, of linear size up to $L=512$, which is crucial for obtaining universal properties. To validate the universality of our results we performed our analysis on two distinct crossing points of the SIT, by choosing $\mu=-0.5$ and $\mu=-5.89391$ and tuning $g$ across the transition.  We found excellent agreement within the error bars.  Henceforth we will only present results for $\mu=-5.89391$, a value which has been argued to reduce finite size corrections to scaling \cite{hasenbusch_high-precision_1999}.


First we locate the critical coupling $g_c(\mu)$ with high accuracy. This can achieved by a finite size scaling analysis of the superfluid stiffness $\rhos=\frac{1}{L}\frac{\partial \mathcal{Z}(\varphi)}{\partial \varphi}|_{\varphi=0}$ \cite{Fisher_Helicity}, where $\mathcal{Z}(\varphi)$ is the partition function in the presence of a uniform phase twist $\varphi$.  In this work we find $g_c=7.0284(3)$.

We extract the gap $\Delta$ in the insulator by analyzing the asymptotic large imaginary time decay of the two point Green's function.  In a gapped phase this has an exponential decay of the form $G(\tau)\sim e^{-\Delta \tau}+e^{-\Delta(\beta-\tau)}$, where $\beta=1/T$. We compute $\Delta$ by a fit to this form. Near criticality, the gap is expected to scale as a power law $\Delta=\Delta_0 \left|\delta g \right|^\nu$, where $\nu$ is the correlation length exponent. We use $\nu=0.6717(3)$, as obtained by previous high accuracy Monte Carlo studies of the 3D XY model \cite{nu_MC}. Our results for $\Delta(\delta g)$ are in excellent agreement with the expected scaling, with the non universal pre-factor $\Delta_0=2.09(5)$.

\begin{figure}
\centering
\includegraphics[scale=.55]{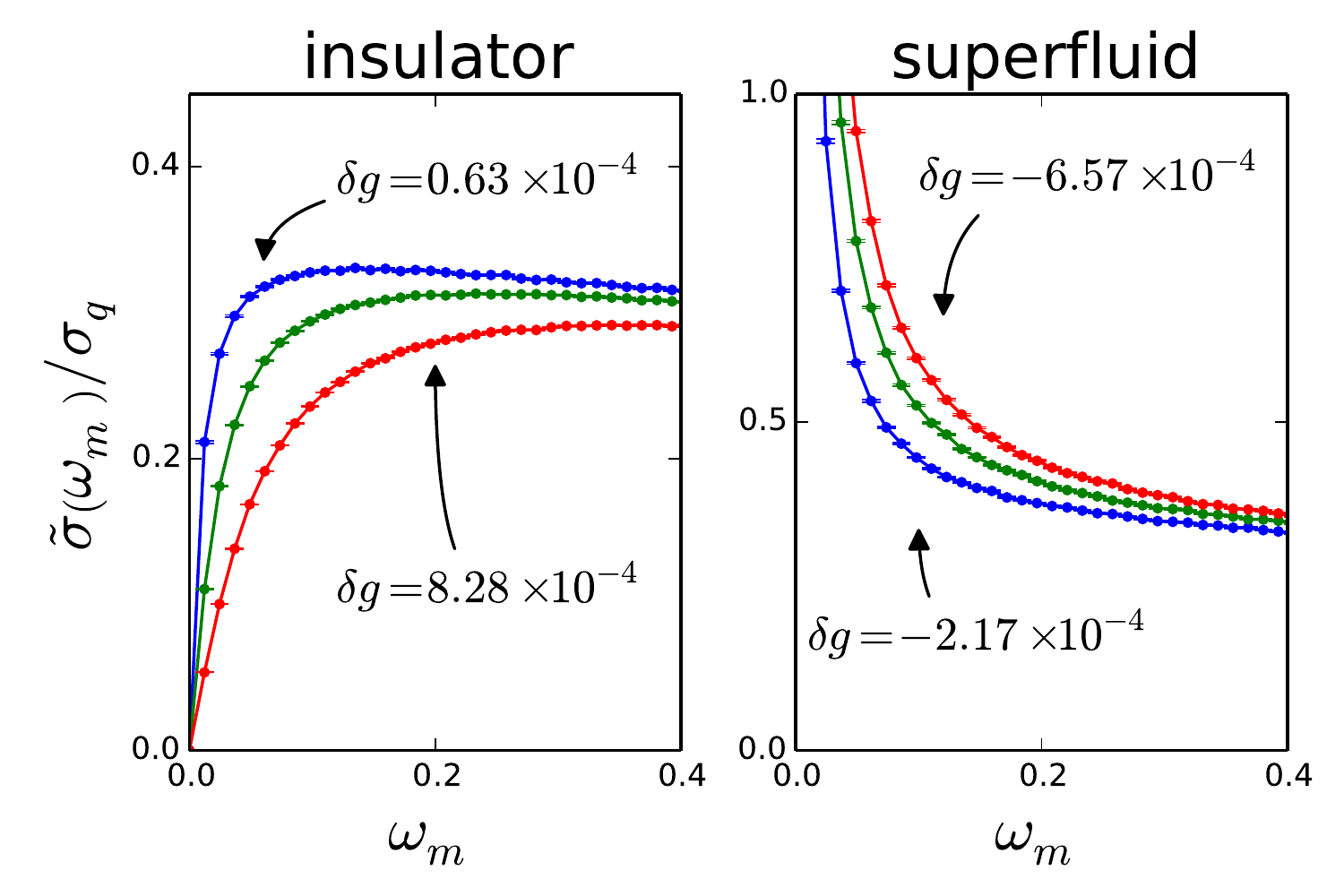}
\caption{The conductivity as a function Matsubara frequency. The curves differ by the detuning parameter $\delta g$. In the insulator, the low frequency conductivity is linear, $\sigma_{\rm ins}\sim \omega_m$ indicating capacitive behavior. In the superfluid, the conductivity diverges as $\sigma_{\rm sf}\sim 1/\omega_m$ indicating inductive response.
}
\label{fig:cond}
\end{figure}

{\em Results --} 
In Fig.~\ref{fig:cond} we present the dynamical conductivity $\sigma(\omega_m)$ as a function of Matsubara frequency, both in the insulator and in the superfluid, for a range of detuning parameters $\delta g$ near the critical point. To suppress finite size effects in the insulator we used an improved estimator, in which we consider only loop configurations with zero winding number \cite{Kun_cond,Witczak_Cond}. We find that the dynamical conductivity as a function of $\omega_m$ in Fig. \ref{fig:cond}, follows the form of the low frequency reactive conductivity both in the superfluid, Eq.~\eqref{eq:sigmasf}, and in the insulator, Eq.~\eqref{eq:sigmains}.

Next we calculate $\rhos$ and $\rhov$ in their respective phases. The superfluid stiffness $\rhos$ was calculated using the standard method of winding number fluctuations \cite{ceperly_wind}. In order to extract $\rhov$ we use the relation in Eq.~\eqref{eq:C_rhov}. As a concrete Monte Carlo observable for the capacitance we use the conductivity evaluated at the first non-zero Matsubara frequency:
\begin{equation}
C({\delta g})=\lim_{L\to \infty}\frac{\sigma(\omega_m=\frac{2\pi}{L},\delta g)}{\frac{2\pi}{L}}.
\end{equation}

Both the vortex condensate stiffness $\rhov$ and the superfluid stiffness $\rhos$ near the critical point follow a power law behavior $\rho_{\{\rm s,v\}} \sim \rho^0_{\{\rm s,v\}}|\delta g|^\nu$. The non-universal prefactors $\rhov^0$ and $\rhos^0$ are extracted by a numerical fit. We find $\rhos/\rhov=0.21(1)$. Finite size scaling effects are discussed in the supplemental material  \cite{supp}.   Surprisingly, this value is close to the value $\rhos/\rhov=0.23$ obtained by a simple one loop weak coupling calculation \cite{Gazit_Dynamics}.

The universal scaling function of the dynamical conductivity is obtained by rescaling the Matsubara frequency axis by the single particle gap $\Delta$. Curves for different detuning parameters $\delta g$ collapse into a single universal curve at low frequencies.   On the other hand, at high frequencies, $\omega_m$ need not be a negligible fraction of the ultra-violet (UV) cutoff scale $\Lambda$.  This leads to non-universal corrections in the conductivity that depend on powers of $\omega_m/\Lambda$.  We take these into account by fitting the numerical QMC data to the following scaling form
\begin{eqnarray}
\sigma_{\pm}(\omega_m,\delta g,\Lambda)=\sigma_q\Sigma_{\pm}\left(\frac{\omega_m}{\Delta}\right)+A\, \frac{\omega_m}{\Lambda}+ B\left(\frac{\omega_m}{\Lambda}\right)^2.
\label{eq:sub_lead_scale}
\end{eqnarray}
Here, $A$ and $B$ are expected to depend smoothly on the detuning parameter $\delta g$.  Since we consider a narrow range of values of $\delta g$, we approximate $A$ and $B$ as constants. This enables us to extract the universal functions $\Sigma_\pm$ on  {\em both phases} by using only two fitting parameters. For further details, see the supplemental material \cite{supp}.

\begin{figure}
\centering
\includegraphics[scale=.5]{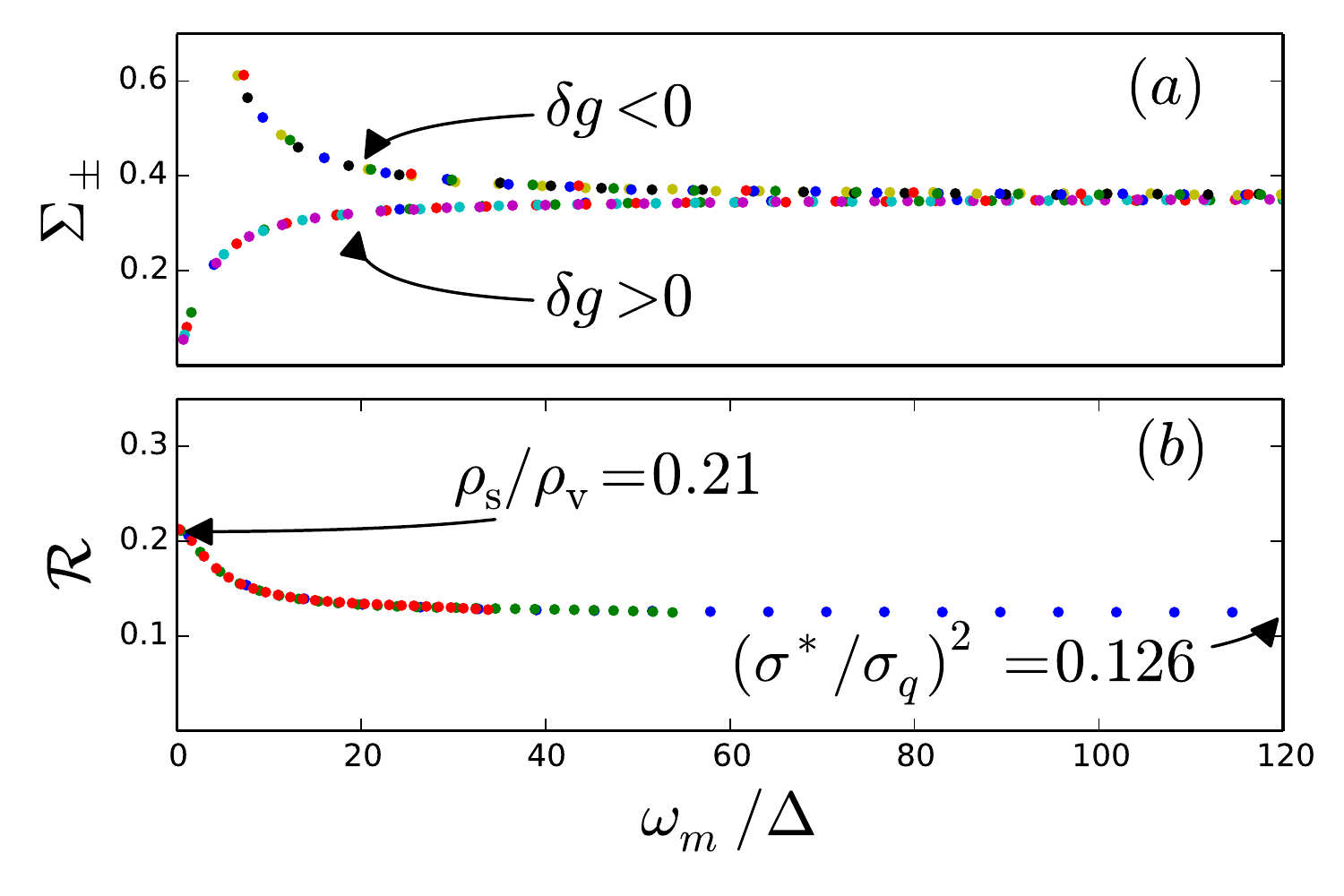}
\caption{(a) Scaling function of the dynamical conductivity in the superfluid ($\delta g<0$) and insulator ($\delta g>0$). Data for different values of the detuning $\delta g$ collapse to two universal curves.  (b)  Measure of charge vortex duality of the $O(2)$ model. Universal scaling function for ${\mathcal{R}}(\omega_m)$ defined in Eq.~\eqref{eq:calR}.  
Deviation of this function from unity quantifies the difference between charge and vortex matter.
}
\label{fig:cond_scaled}
\end{figure}
The result of this analysis is shown in Fig. \ref{fig:cond_scaled}(a), where we subtract the non universal part of the conductivity using Eq. \eqref{eq:sub_lead_scale}. The conductivity curves, on each side of the phase transition, collapse, with high accuracy, to the universal conductivity functions $\Sigma_{\pm}(\omega_m/\Delta)$.

At high frequencies the universal conductivity curves saturate to a plateau, with  $\sigma(\omega\gg\Delta)=0.354(5)\, \sigma_q$ in the insulating phase and $\sigma(\omega\gg\Delta)=0.355(5)\, \sigma_q$ in the superfluid phase. As a result, we conclude that the high frequency universal conductivity, $\sigma^*$, is a robust quantity across the phase transition.  Our scaling correction analysis differs significantly from that of Refs.~\cite{Witczak_Cond,Kun_cond,holo_sachdev}, yet the value of the high-frequency conductivity is in agreement with their results.

Finally, we study deviations from self-duality as a function of Matsubara frequency. In Fig.~\ref{fig:cond_scaled}(b) we depict the product of the Matsubara frequency conductivity evaluated at mirror points across the critical point, ${\mathcal{R}}(\omega_m)=\sigma(\omega_m,\delta g)\sigma(\omega_m,-\delta g)/\sigma_q^2$. In order to study the critical properties we subtract the non-universal cut-off corrections. 
Note that for $\omega_m\gg \Delta$, ${\mathcal{R}}\to(\sigma^*/\sigma_q)^2$, whereas for $\omega_m\ll \Delta$, ${\mathcal{R}}$ approaches the product of reactive conductivities in the two phases.  In both limits, the Matsubara and real frequency results coincide, ${\mathcal{R}}(\omega)={\mathcal{R}}(\omega_m)$.  By contrast, at intermediate frequencies, determination of ${\mathcal{R}}(\omega)$ requires analytical continuation.
If the CVD were self-dual then Eq.~\eqref{eq:recip} would imply that this product is frequency independent and equal to $1$. Our results display a non trivial frequency dependence and deviate from the predicted self-dual value. We attribute this deviation to the different interaction range of charges and vortices.

{\em Discussion and summary - } The universal ratio of the reactive conductivity $C_{\rm ins}/L_{\rm sf}$ motivates future experiments as it provides a direct probe of the charge-vortex duality. 

Recent THz spectroscopy measured the complex AC conductivity near the SIT in superconducting InO and NbN thin films \cite{shermanThz}. In these systems, the superfluid stiffness in the superconducting phase can be measured from the low frequency reactive response \cite{corson_vanishing_1999,crane_fluctuations_2007}. Detecting the diverging capacitance in the insulating side may require careful subtraction of substrate signal background \cite{private_fry}.

Another experimental realization of the SIT is the Mott insulator to superfluid transition of cold atoms trapped in an optical lattice.  At integer filling, the transition has an emergent Lorentz invariance \cite{fisher_boson_1989} and hence its critical properties are captured by the $O(2)$ relativistic model in Eq.~\eqref{eq:lat_model}. We propose a direct approach to extract the capacitance of the Mott insulator using {\em static} measurements.  In the insulator, the current and charge response functions are related by the continuity equation, $\Pi_{xx}(k,\omega)=-\frac{\omega^2}{k^2}\chi_{\rho}(k,\omega)$, where $\chi_{\rho}(k,\omega)$ is the charge susceptibility.  Hence,
\begin{eqnarray}
C_{\rm ins}=\lim_{\omega\to 0} \lim_{k\to 0}\frac{\Pi_{xx}(k,\omega)}{-\omega^2}=\lim_{k\to 0}\frac{\chi_\rho(k,\omega=0)}{k^2}\label{eq:compr}\,,
\end{eqnarray}
where the  $\omega\to 0$ and $k\to 0$ limits commute  since the insulator is gapped \cite{Scalapino_criteria}.   Thus, the capacitance is simply related to the finite $k$ compressibility of the Mott insulator.  This can be measured, {\em e.g.} by applying an optical potential at small wave vector $k$ and probing the rearrangement of boson density using {\em in-situ} imaging \cite{insitu}.  Temperature effects are discussed in the supplemental material  \cite{supp}.

Alternatively $\sigma'(\omega)$, for which experimental protocols were proposed,  \cite{Giamarchi_Phase,Kun_cond} can be used to compute $\sigma''(\omega)$ by means of the Kramers-Kronig integral.

In summary, we computed the vortex condensate stiffness $\rhov$, the high frequency universal conductivity, and provided a quantitative measure for deviation from self-duality as a function of Matsubara frequency. In addition, we suggest concrete experiments that test our predictions in THz spectroscopy of thin superconducting films and in cold atom systems.

{\em Acknowledgements.--} We thank D.~P.~Arovas, M.~Endres, A.~Frydman, and W. ~Witczak-Krempa for helpful discussions.
AA and DP acknowledge  support from the Israel Science Foundation, the European Union under grant IRG-276923, the U.S.-Israel Binational Science Foundation.  We  thank the Aspen Center for Physics,  supported by the NSF-PHY-1066293, for its hospitality. SG acknowledges the Clore foundation for a fellowship.
\appendix

\onecolumngrid

\parindent=0pt

\graphicspath{{./}}
\pagebreak
\section{\large Supplementary Material for ``Critical capacitance and charge-vortex duality near the superfluid to insulator transition"}
In the supplemental materials below, we explain the high frequency correction to scaling in Eq. (12) of the main text. In addition, we  describe finite temperature effects on the capacitance in the insulating phase, which may prove useful for
analysis of experimental data.
\maketitle In the supplemental materials below, we present a finite size scaling analysis of the critical energy scales, we explain the high frequency correction to scaling in Eq. (12) of the main text and we describe finite temperature effects on the capacitance in the insulating phase, which may prove useful for
analysis of experimental data.
\subsection{Finite size scaling analysis of the critical energy scales }

In this section we present the numerical protocol used to compute the critical energy scales near the QCP. In particular, we demonstrate power law scaling and discuss finite size scaling effects. We consider three critical energy scales: the superfluid stiffness $\rho_{\rm s}$, obtained from the winding number statistics \cite{ceperly_wind}; the vortex condensate stiffness $\rho_{\rm v}$, evaluated according to Eq. (11) in the main text; and the single particle gap $\Delta$, computed by a fit to the asymptotic exponential decay of the two point Green's function as a function of imaginary time \cite{Gap2DBose}.  Close to the QCP,  we expect the critical energy scales to follow the scaling behavior,
\begin{eqnarray}
E_\alpha=E_\alpha^0 |\delta g|^\nu F_\alpha(|\delta g|^{-\nu}/L)
\label{eq:finiteSize}
\end{eqnarray}
where  $\nu=0.6701$ is the correlation length exponent  for the 2+1 dimensional XY universality class, and $F_\alpha(x)$ is a finite size scaling function that tends to 1 in the thermodynamic limit $x=0$.  More generally, the function $F_\alpha$ takes into account corrections that arise when the correlation length is not negligible compared to the system size. The subscript $\alpha=1$, $2$, and $3$ is used to label $\rho_{\rm s}$, $\rho_{\rm v},$  and $\Delta$, respectively.
\begin{figure}[h]
	\centering
	\includegraphics[scale=.4]{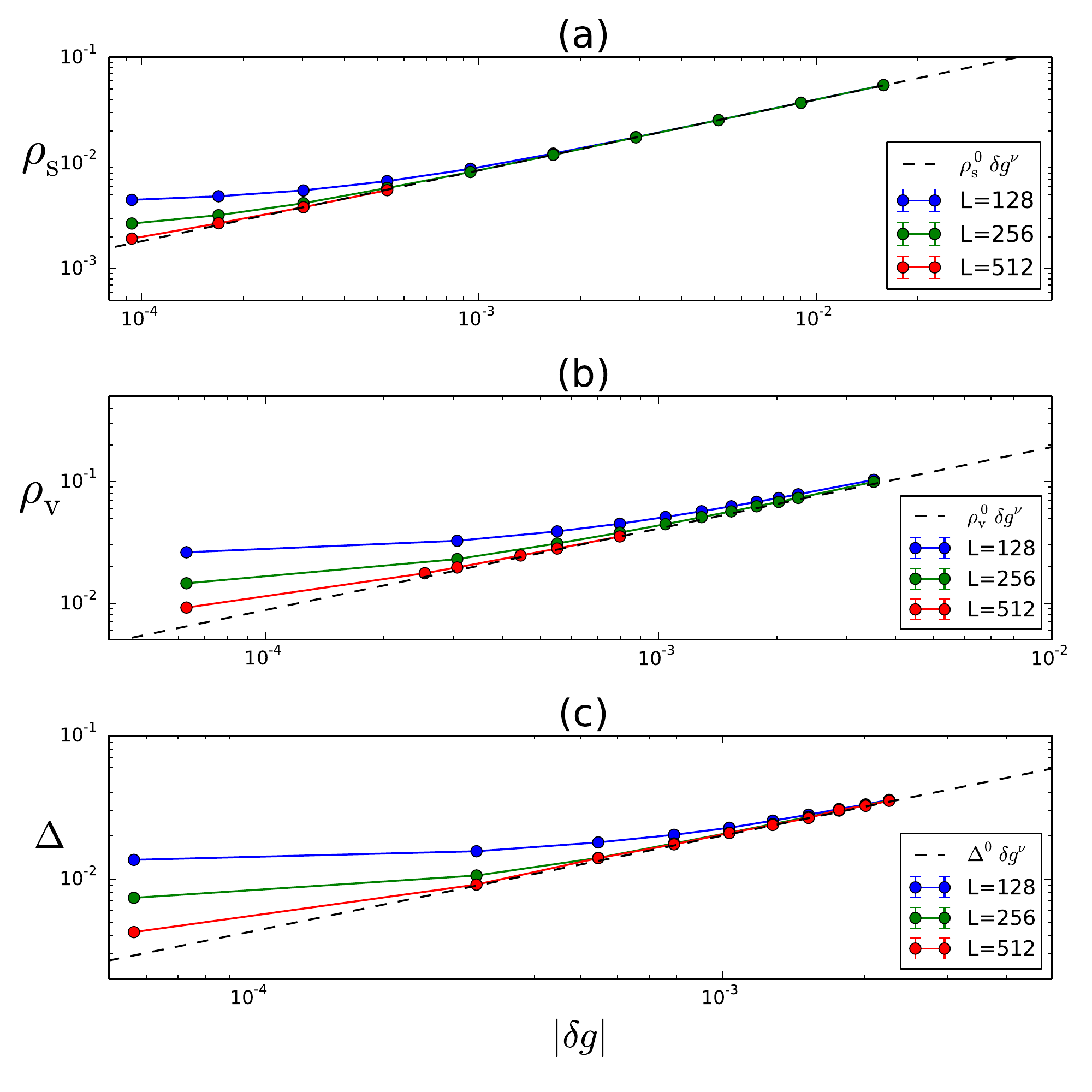}
	\caption{ Critical energy scales in the QCP as a function of the detuning parameter $\delta g$ and linear system size $L$ displayed in a log-log plot. The different panels contain, (a) the superfluid stiffness $\rho_{\rm s}$ (b) the vortex condensate stiffness $\rho_{\rm v}$ and (c) the single particle gap $\Delta$. All energy scales follow the expected power scaling. 
	}
	\label{fig:ffs}
\end{figure}

Figure \ref{fig:ffs} depicts the critical energy scales on a log-log plot for a range of  detuning parameters $\delta g$ in close proximity to the QCP, and for a range of linear system sizes $L$. In all cases, we observe deviations from the pure power law scaling behavior $E_\alpha=E_\alpha^0 |\delta g|^\nu$.  These deviations occur very close to the QCP, where the correlation length is largest, but for any fixed detuning $\delta g$ they are seen to vanish as $L\to\infty$, as expected from Eq.~(\ref{eq:finiteSize}). The amplitudes $E_\alpha^0$ are then extracted by a numerical fit. The result of this analysis is depicted by dashed lines in Fig.~\ref{fig:ffs}.

\begin{figure}[h]
	\centering
	\includegraphics[scale=.4]{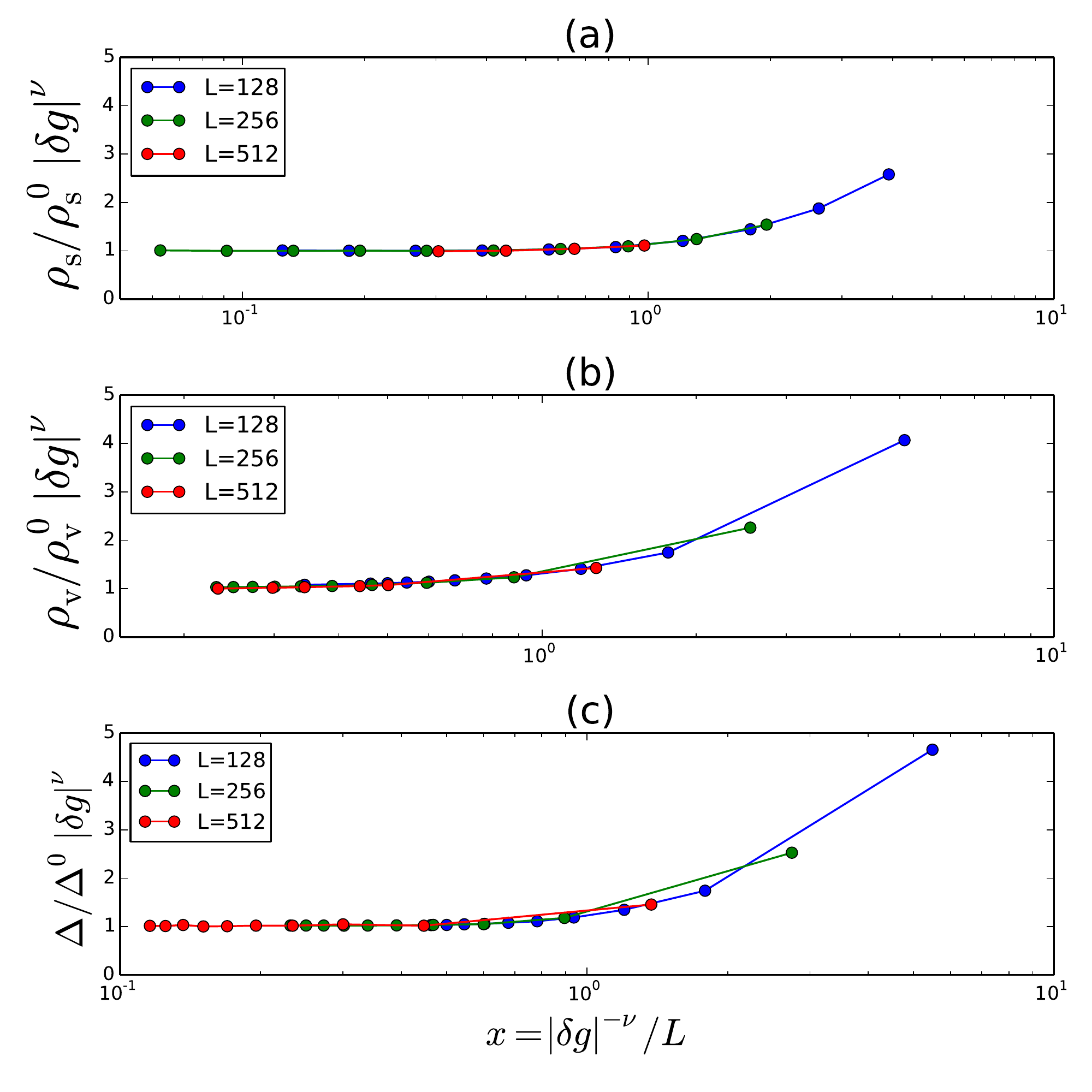}
	\caption{ Finite size scaling functions $F_\alpha(x)$, see Eq.~\eqref{eq:finiteSize}. The energy axis is rescaled by $E_\alpha^0 |\delta g|^\nu$ and it is plotted as a function of $x=|\delta g|^{-\nu}/L$.
	}
	\label{fig:ffs_function}
\end{figure}

We further analyze the finite size effects by computing the scaling functions $F_\alpha(x)$  defined in Eq.~\eqref{eq:finiteSize}. This is accomplished by performing a joint rescaling of the critical energy, $E_\alpha$ and the coupling detuning, $\delta g$, according to the scaling relation in Eq.~\eqref{eq:finiteSize}. The result of this analysis is presented in Fig.~\ref{fig:ffs_function}. We find that the critical energy curves, associated with different linear system sizes $L$, collapse into a single universal curve. Importantly, the scaling functions rapidly converge to the thermodynamic value $F_\alpha(x\to0)=1$.

\subsection{High frequency nonuniversal corrections to the dynamical conductivity  }
In this section we analyze, in detail, the high frequency nonuniversal corrections to the dynamical conductivity. The conductivity is a universal amplitude, as a result, the universal scaling function is obtained by rescaling the frequency axis $\omega_m$ by the critical energy scale $\Delta$. At low frequency, curves for different detuning parameters $\delta g$  collapse into a single universal curve.  At high frequency, closer  to the ultraviolet cutoff scale $\Lambda$, one expects  a substantial deviation from scaling.

\begin{figure}
	\subcaptionbox{\label{subfig:unscaled}}
	{
		\includegraphics[scale=.5]{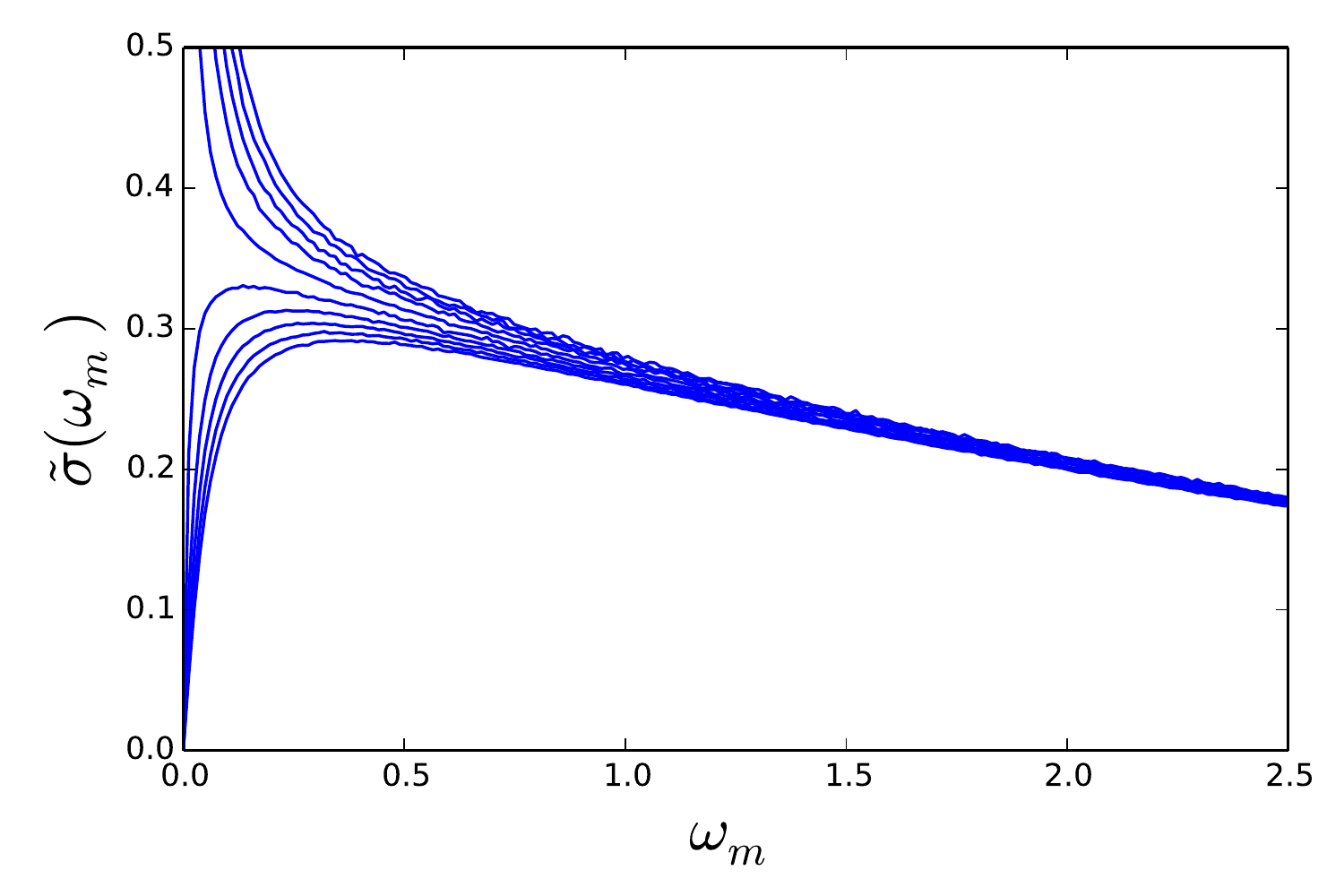}
	}
	\subcaptionbox{\label{subfig:scaled}}
	{
		\includegraphics[scale=.5]{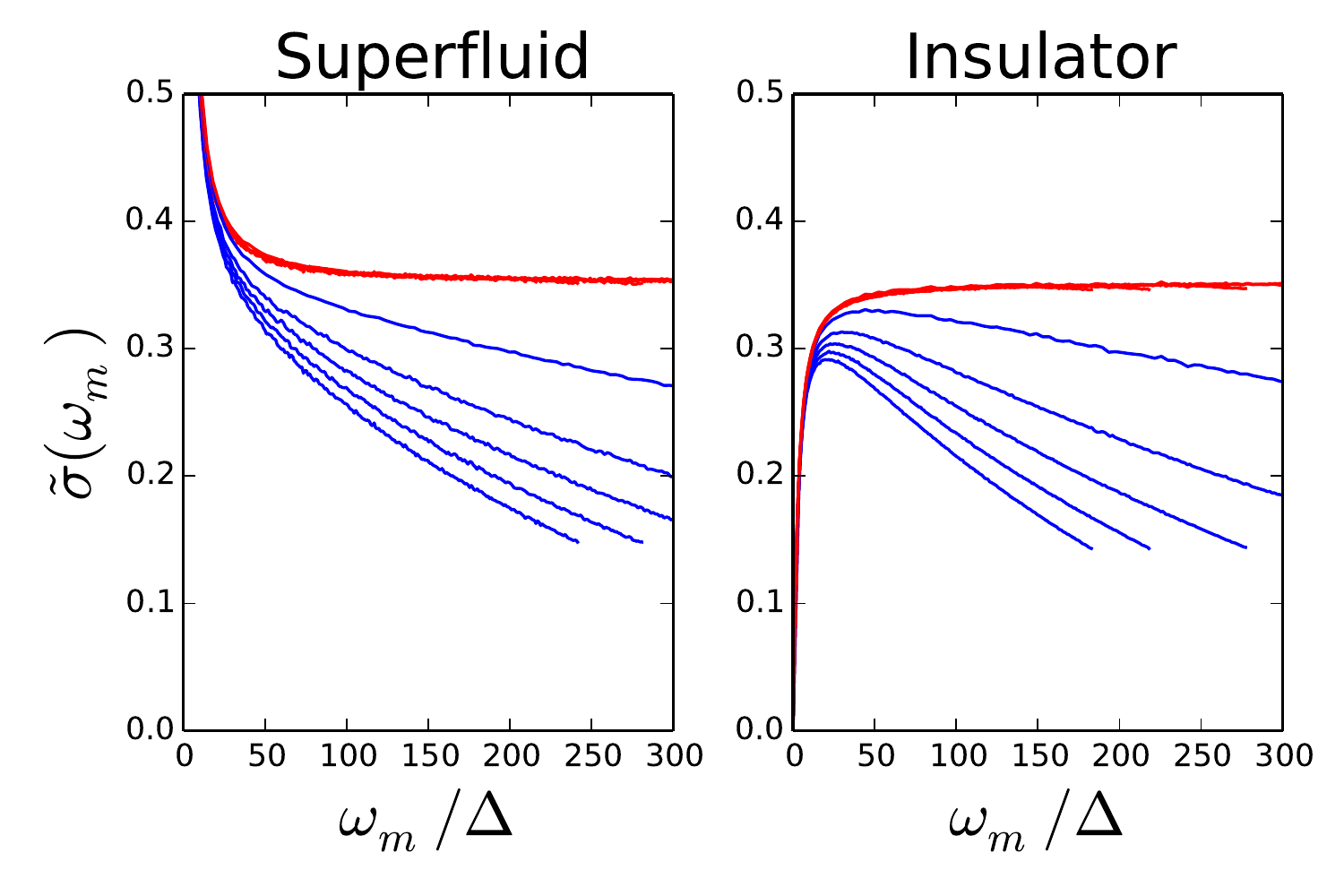}
	}
	
	\caption{Matsubara conductivity as a function of (a)  frequency $\omega_m$ and (b) rescaled  frequency $\omega_m/\Delta$, for a range of detuning parameters $\delta g$ near the QCP both in the superfluid and the insulator. Curves before (after) subtraction of the high frequency corrections to scaling are shown in blue (red).}
	\label{fig:corr_scal}
\end{figure}

Figure~\ref{subfig:unscaled} shows the Matsubara conductivity for a range of values of $\delta g$ on either side of the SIT.  The cutoff in this plot is $\Lambda=\pi$,  the maximal value of the Matsubara frequency.  For $\omega_m\ll \Lambda$, the conductivity depends sensitively on $\delta g$.  By contrast,  when $\omega_m/\Lambda$ is no longer negligible, {\em e.g.} for $\omega_m>1$, the conductivity evolves smoothly across the transition.  This is expected,  since correlations at large frequencies and short time scales are insensitive to the phase transition, and are instead functions that depend on the details of the model at the lattice scale.

In order to eliminate these nonuniversal corrections we consider the following ansatz for the scaling function near the SIT,
\begin{equation}
\sigma(\omega_m,\Delta,\delta g)=\sigma_q \Sigma_\pm(\omega_m/\Delta)+\sigma_{\rm nu}(\omega_m/\Lambda,\delta g)\, ,
\end{equation}
where $\sigma_{\rm nu}$ is a nonuniversal function that depends smoothly on $\delta g$, and that vanishes when $\omega_m/\Lambda\to 0$. Expanding $\sigma_{\rm nu}$ in powers of $\omega_m/\Lambda$ gives,
\begin{equation}
\sigma(\omega_m,\Delta,\delta g)\approx\sigma_q\Sigma_\pm(\omega_m/\Delta)+A\omega_m +B\omega_m^2\, .
\label{eq:scaling_form}
\end{equation}
Here we have absorbed the $\Lambda$ dependence into the nonuniversal coefficients $A$ and $B$, which depend smoothly on the detuning parameter $\delta g$.  Since we consider a narrow range of values of $\delta g$, we take $A$ and $B$ as constants. The high frequency Matsubara conductivity $\tilde{\sigma}(\omega_m\gg \Delta)$ curves, shown in Fig.~\ref{subfig:unscaled}, display a clear linear decrease as a function of Matsubara frequency. Importantly, all curves on {\em both} phases share the {\em same} linear slope (and the same weak upward curvature). This behavior exactly matches the correction to scaling ansatz predicted in Eq.~\eqref{eq:scaling_form}.

\begin{figure}[h]
	\centering
	\includegraphics[scale=.5]{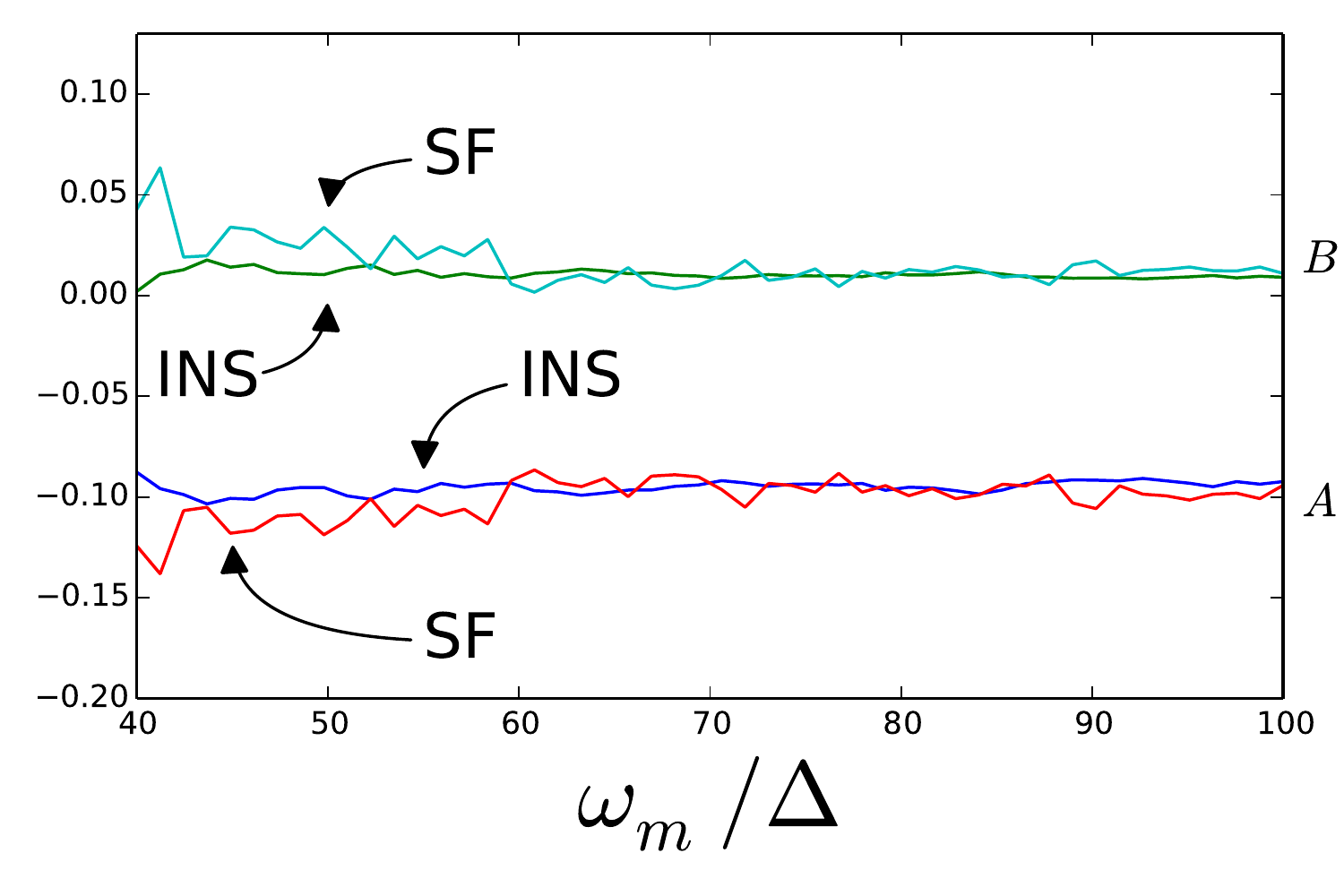}
	\caption{Numerical fit of the parameters $A$ and $B$ in Eq.~\eqref{eq:scaling_form}, as a function of $\omega_m/\Delta$.  Both quantities show little variation with respect to $\Delta$ or $\delta g$. Importantly, they vary smoothly across the phase transition, as the values of $A$ and $B$ do not change significantly from the insulating (INS) phase to the superfluid (SF). 
	}
	\label{fig:AB_Fit}
\end{figure}

To verify our ansatz we determined the parameters $A$ and $B$ by a numerical fit to Eq.~\eqref{eq:scaling_form} at fixed values of $\omega_m / \Delta$ on both sides of the phase transition. The results of this analysis are presented in Fig.~\ref{fig:AB_Fit}. We find, within the noise in the fit, that both $A$ and $B$ are {\em independent} of $(\omega_m/\Delta)$ and do not change when crossing the SIT. This provides strong evidence for validity of our ansatz and demonstrates that the correction to scaling originates from the short range physics at the cutoff scale $\Lambda$ and not from the correction to the critical energy scale $\Delta$.  To increase the accuracy of our fit we performed a joint fit using all frequency  and $\delta g$ data points. The resulting values of $A$ and $B$ were used in the main text.

In Fig.~\ref{subfig:scaled} we depict the Matsubara conductivity as a function of rescaled frequency $\omega_m/\Delta$ for a range of detuning parameters $\delta g$ near the QCP both in the superfluid and in the insulator. We then subtract the nonuniversal high frequency corrections according to Eq.~\eqref{eq:scaling_form} and display the resulting curves. The rescaled curves collapse, with high accuracy, to the universal scaling functions corresponding to the superfluid and insulator.  
We emphasize that the same values of $A$ and $B$ are used for all curves. 

It is interesting to compare our analysis with the one presented in Ref.~\cite{Witczak_Cond,Kun_cond,holo_sachdev}, which considers corrections to scaling that arise from irrelevant renormalization group (RG) eigenvalues. The foregoing studies focused on finite temperature scaling at the critical regime thus the temperature $T$ plays the role of the critical energy scale. The leading correction to the scaling form is then \cite{cardy1996scaling}
\begin{equation}
\sigma(\omega_m/T)=\sigma_q \Sigma(\omega_m/T)+f(\omega_m/T)T^{\varpi}.
\label{eq:scaling_temp}
\end{equation} 
In the above equation the critical exponent $\varpi$ is the smallest irrelevant RG eigenvalue. The authors of Ref.~\cite{Kun_cond} take $\varpi=0.85$ based on previous studies of the three dimensional XY universality class. In addition, they observe that the nonuniversal prefactor follows the functional form $f(\omega_m/T)=A_0+A_1(\omega_m/T)^{\varpi}$ with the same exponent $\varpi$ as before. Both $A_0$ and $A_1$ are taken to be constant fitting parameters. The exact values of $A_0$ and $A_1$ are not specified in the manuscript, but one can clearly deduce from Fig 4. of the supplementary material section of \cite{Kun_cond} that $A_0\approx0$. In this case the temperature dependence of the second term in Eq.\eqref{eq:scaling_temp} vanishes, such that
\begin{equation}
\sigma(\omega_m/T)=\sigma_q \Sigma(\omega_m/T)+A_1\omega_m^{\varpi}
\end{equation}        
The authors of Ref.~\cite{Kun_cond} further claim that when $\varpi$ is considered as a free fitting parameter then values in the range $\varpi=0.9\pm0.1$ produce fits which are consistent with the error bars. We note that for $\varpi=1$ both methods use essentially the {\em same} ansatz for corrections to scaling and hence our results are consistent with the one obtained in \cite{Witczak_Cond,Kun_cond,holo_sachdev}. It is important to notice that even though the two method use a similar mathematical expression in the fitting procedure, their underling reasoning is distinct. We found that the corrections to scaling do not depend on the critical energy scale $\Delta$ and they are sizable only at high frequency. We believe that this provides strong evidence that the corrections to scaling originate from high frequency (cutoff scale) effects. It would be interesting to apply our approach to finite temperature scaling in future research.  

\subsection{Charge susceptibility at finite temperature}

In this section we study the charge susceptibility $\chi_{\rho}(k,\delta g,T)$ at finite temperature. 
The charge operator in the O(2) model is defined as $\rho=i\left(\psi^*\partial_t\psi - \psi\partial_t\psi^*\right)$.

\begin{figure}[h]
	\includegraphics[scale=.5]{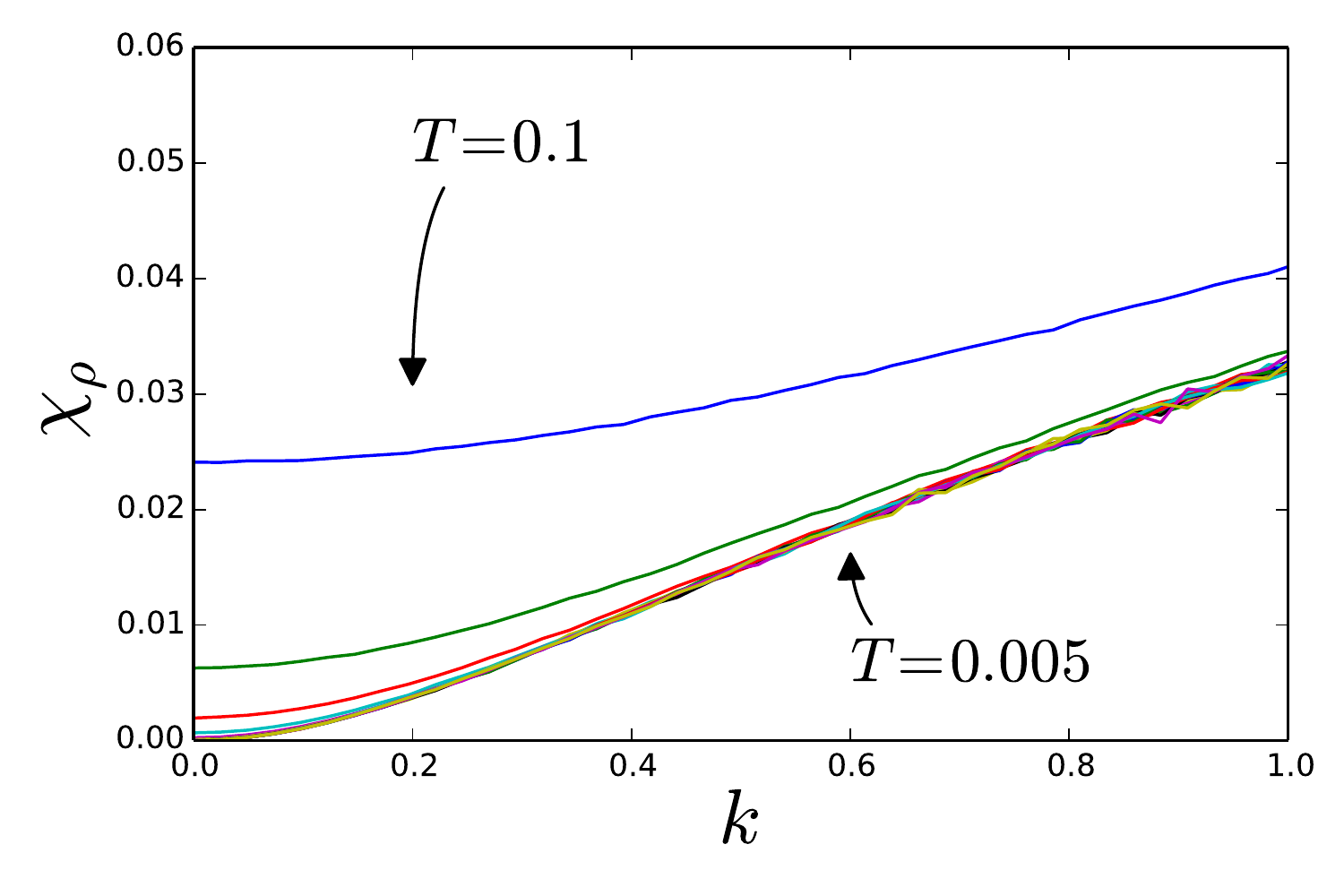}
	\caption{The charge susceptibility $\chi_{\rho}(k,T)$ as a function of momentum $k$. We fix $\delta g=1.12\times10^{-2}$, for which $\Delta=0.103$. Each curve corresponds to a different temperature $T$. The temperature T and $\Delta$ are measured in units of the inverse lattice constant.}
	\label{fig:chi_q_t}
\end{figure}

As an illustrative example, in Fig.~\ref{fig:chi_q_t} we depict $\chi_{\rho}(k,T)$, obtained from MC simulations of Eq.~(5) in the main text, for a fixed detuning parameter $\delta g=1.12\times10^{-2}$ and a range of temperatures $T$. At finite temperatures the compressibility $\chi_\rho^0=\chi_\rho(k=0,T)$ is non-zero. In addition, the curvature at zero momentum, $\frac{d^2\chi_\rho(k,T)}{dk^2}|_{k=0}$, decreases at finite temperatures. Both effects should be taken into account in measurements of the capacitance at finite temperature.

\begin{figure}[h]
	\includegraphics[scale=.5]{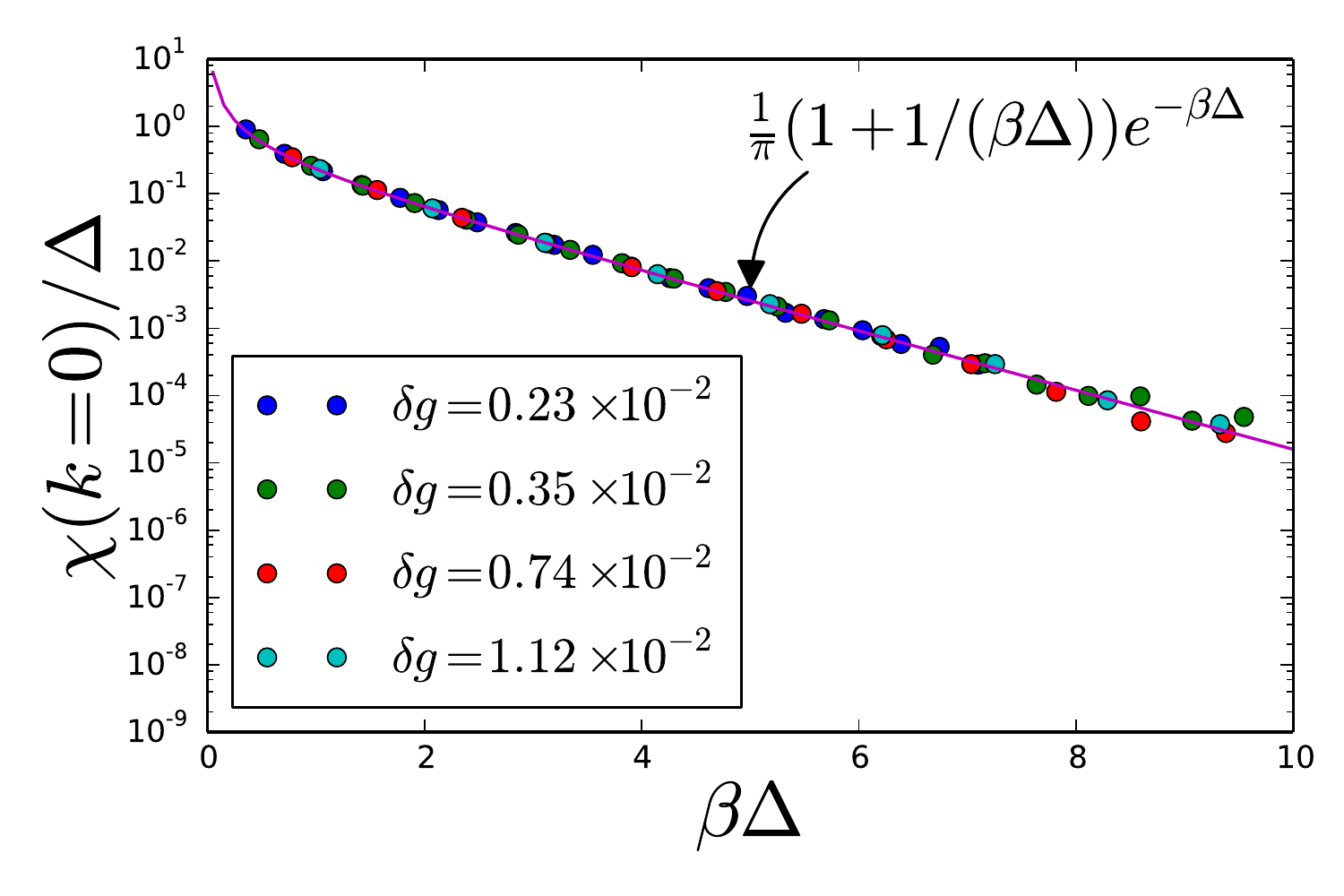}
	\caption{Scaling function of the compressibility $\chi_{\rho}(k=0,\delta g,T)=\Delta\, f(\beta\Delta)$.  Curves for different values of the detuning parameter $\delta g$ collapse to a single universal curve. The functional form of the curve matches the analytic calculation for a free Klein-Gordon field. }
	\label{fig:chi_zero_scal}
\end{figure}

The insulator is a gapped phase, therefore we expect that at low temperatures the compressibility will follow an activated behavior \cite{endres_thesis}. This is demonstrated in Fig.~\ref{fig:chi_zero_scal}, where we plot the compressibility as a function of the inverse temperature $\beta$. The compressibility scales near the critical point as  $\chi_\rho(k=0,\beta,\delta g)=\Delta\, f(\beta\Delta)$, and both axes in Fig.~\ref{fig:chi_zero_scal} are rescaled to obtain the collapsed universal scaling function $f$. Indeed, we find that at low temperatures, $\beta\Delta \gg 1$, the compressibility is activated,  $\chi_\rho^0\sim e^{-\Delta/T}$. 

\begin{figure}[h]
	\includegraphics[scale=.5]{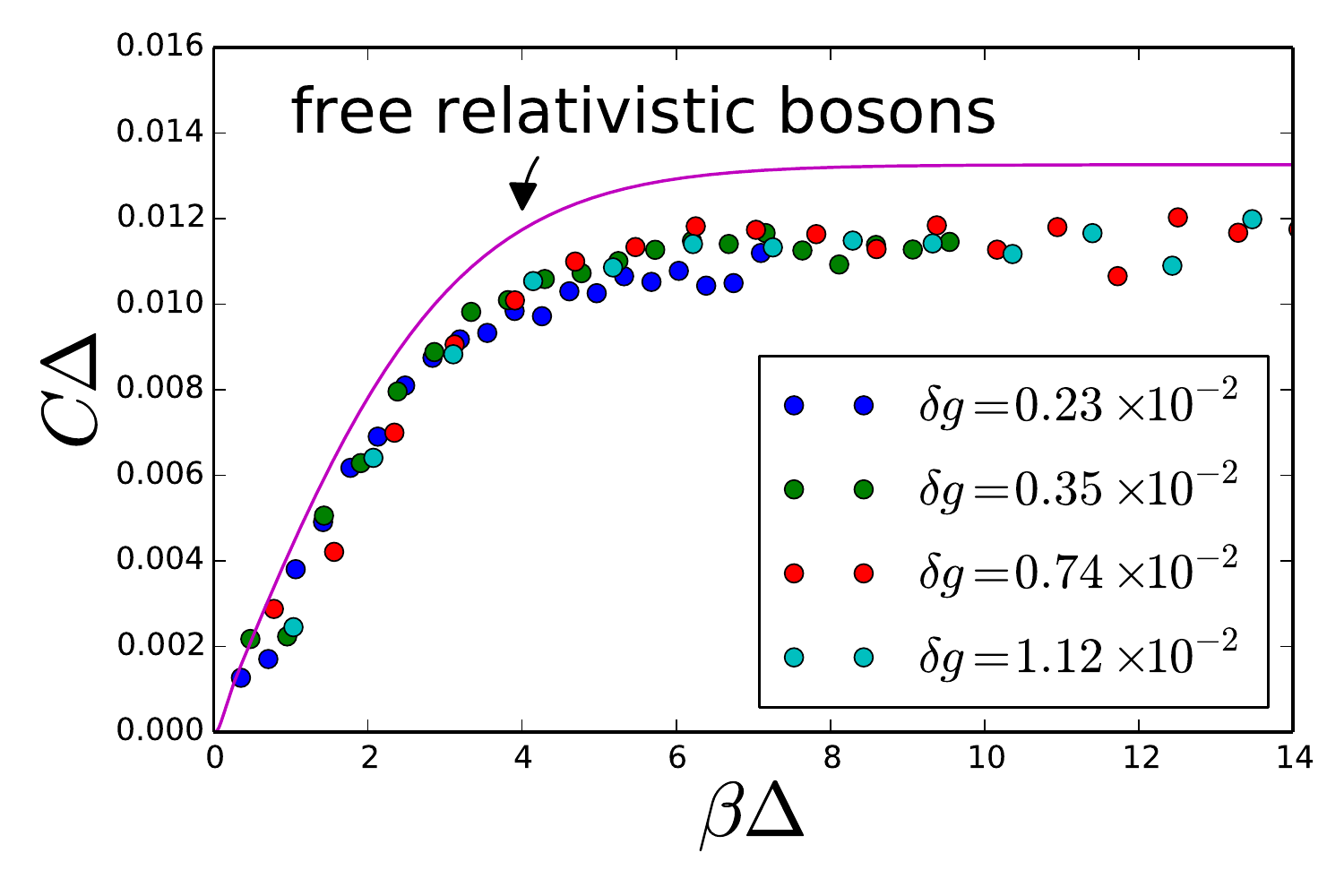}
	\caption{Curvature of the charge susceptibility, $C(T)=\frac{1}{2}\frac{\partial^2\chi_\rho(k)}{\partial k^2}|_{k=0}$.  For $T\to 0$, this becomes the capacitance.  Axes are rescaled to obtain scaling behavior.  The noise is dominated by numerical derivatives of the QMC data.  The solid line shows an analytic calculation of $C$ for a free Klein-Gordon field.}
	\label{fig:C_T_scale}
\end{figure}

We define a generalization of the capacitance to finite temperatures as the curvature of charge susceptibility at zero wave number, 
\begin{eqnarray}
C(T)=\frac{1}{2}\frac{\partial^2\chi_\rho(k)}{\partial k^2}|_{k=0}\,.
\label{eq:CT}
\end{eqnarray}
For $T=0$ this coincides with the zero temperature definition. In Fig.~\ref{fig:C_T_scale} we depict the generalized capacitance as a function of the inverse temperature $\beta$.  As before, both axes are rescaled to obtain the scaling function. The curves rapidly converge to the zero temperature limit, allowing for an accurate determination of the capacitance for $\beta \Delta >6$.  

As a point of reference, we compare our compressibility to that of a simple analytic calculation based on a free complex Klein-Gordon boson field with mass $\Delta$, with Euclidean action
\begin{eqnarray}
S=\int_0^\beta d\tau \int d^2 x\left[\left|\partial_\tau\psi\right|^2+ \left|\nabla\psi\right|^2+\Delta^2\left|\psi\right|^2\right]\,.
\end{eqnarray}
This gives the leading result in a $1/N$ expansion, using the renormalized mass $\Delta$ as an input.
The static charge susceptibility is then obtained from a one-loop Feynman diagram calculation, \cite{damle_UniCond}
\begin{equation}
\chi_\rho(k,\nu_m=0,T)=\frac{1}{\beta}\sum_{\omega_m}\int \frac{d^2p}{(2\pi)^2} \left[\frac{4\omega_m^2 }{\left(\omega_m^2+p^2+\Delta^2\right)\left(\omega_m^2+(k+p)^2+\Delta^2\right)}-\frac{2}{\omega_m^2+p^2+\Delta^2}\right]\, ,
\label{eq:oneloop}
\end{equation}
where the second term is the diamagnetic contribution.  At $p=0$ this gives the temperature dependent compressibility,
\begin{equation}
\chi_\rho(k=0,\nu_m=0,T)=\int_0^\infty \frac{p\,d p}{2\pi}\, \frac{\beta}{\cosh (\beta \omega_p)-1}\sim \frac{\Delta}{\pi}\left(1+1/(\beta\Delta)\right)e^{-\beta\Delta}\, .
\end{equation}
where $\omega_p=\sqrt{p^2+\Delta^2}$, and where the last expression is asymptotic in the limit $\beta\Delta\gg 1$.  This expression is shown in Fig.~\ref{fig:chi_zero_scal} to match closely the numerical data, despite the simplicity of the model.
In addition, from Eq.~(\ref{eq:oneloop}) we compute the finite temperature curvature of the charge susceptibility, Eq.~(\ref{eq:CT}).  We obtain the integral expression
\begin{eqnarray}
\frac{1}{2}\left.\frac{\partial^2\chi_\rho}{\partial k^2}\right|_{k=0}&=&\int_0^\infty  \frac{p\,dp}{96 \pi  \omega _p^5} \left\{\beta  \omega _p \text{csch}^2\left(\frac{\beta  \omega _p}{2}\right) \left[6 \Delta ^2+\beta^2p^2  \omega _p^2  \left(3 \text{csch}^2\left(\frac{\beta  \omega _p}{2}\right)\nonumber+2\right)-6\beta  \omega _p^3 \coth \left(\frac{\beta  \omega _p}{2}\right)\right]\right.\\
&\,&\qquad\qquad\qquad+\left.12 \Delta ^2 \coth \left(\frac{\beta  \omega_p}{2}\right)\right\}\end{eqnarray}
where $\text{csch}\, x\equiv 1/\sinh x$.  This is evaluated numerically in Fig.~\ref{fig:C_T_scale}.  In this case, the analytic result captures the qualitative temperature dependence, but it does not yield the correct overall scale.

\bibliography{universal_reactive_cond}{}
\end{document}